# Integrated Localization and Path Planning for an Ocean Exploring Team of Autonomous Underwater Vehicles with Consensus Graph Model Predictive Control

Mohsen Eskandari, Andrey V. Savkin, Mohammad Deghat

*Abstract*—Navigation of a team of autonomous underwater vehicles (AUVs) coordinated by an unmanned surface vehicle (USV) is efficient and reliable for deep ocean exploration. AUVs depart from and return to the USV after collaborative navigation, data collection, and ocean exploration missions. Efficient path planning and accurate localization are essential, the latter of which is critical due to the lack of global localization signals and poor radio frequency (RF) communication in deep waters. Inertial navigation and acoustic communication are common solutions for localization. However, the former is subject to odometry drifts, and the latter is limited to short distances. This paper proposes a systematic approach for localization-aware energy-efficient collision-free path planning for a USV-AUVs team. Path planning is formulated as finite receding horizon model predictive control (MPC) optimization. A dynamic-aware linear kinodynamic motion equation is developed. The mathematical formulation for the MPC optimization is effectively developed where localization is integrated as consensus graph optimization among AUV nodes. Edges in the optimized AUV-to-USV (A2U) and AUV-to-AUV (A2A) graphs are constrained to the sonar range of acoustic modems. The time complexity of the consensus MPC optimization problem is analyzed, revealing a nonconvex NP-hard problem, which is solved using sequential convex programming. Numerical simulation results are provided to evaluate the proposed method.

## I. INTRODUCTION

OCEAN SCIENCE has been facing problems in exploring unvisited underwater environments due to human survival limitations. Besides, immature technologies of underwater vehicles impose physical constraints on the maximum distances these vehicles could reach [1]. Recently, significant advancements have been achieved in marine vehicles regarding navigation, control, sensors, and data collection that enable them to autonomously accomplish underwater tasks [2]. These tasks are often executed using pre-programmed algorithms such as the lawnmower pattern for navy inspection, harbor patrolling, and marine resource detection [3]. Deep ocean exploration is a relatively recent human activity for scientific and commercial purposes, also critical for rescue operations, e.g., deep ocean searches for the missing Malaysia Airlines Boeing 777 [4].

Nevertheless, deep ocean exploration of AUVs faces further challenges such as the absence of light and localization signals [5]. Accurate localization is essential for collecting reliable data and preventing AUVs from becoming lost. However, the global positioning system (GPS) and radio frequency (RF) communication are not available in deep water, which makes localization a particularly difficult task in such dark environments [6]. A primary method for navigation and localization is dead reckoning through the inertial navigation system (INS) (including gyroscopes and accelerometer) and acoustic communication signals [7]. However, the former is subject to odometry drift, and the latter is limited by the range of sonar sensors and acoustic modems [1]. These issues significantly limit the exploration performance and quality. Despite developments in sensor technologies such as pressure sensors for depth detection, magnetic compasses for heading estimation, and Doppler velocity logs (DVLs) for 3D velocity vector extraction, a margin of inaccuracy persists, which becomes large over long distances [8]. Nevertheless, the INS odometry drift can be corrected using beacon-aided positioning. For instance, by utilizing an ultra-short baseline beacon system, the AUV can estimate its position using the phase difference between the signals coming from different beacons [9].

It is more efficient and reliable to employ a team of collaborative AUVs for deep ocean exploration, coordinated by an unmanned surface vehicle (USV) [10]. The USV can act as a beacon due to its ability to determine its accurate position through GPS. However, this requires the AUVs to stay close to the USV, which restricts their exploration range. Alternatively, the AUVs can periodically come close to the USV to update their accurate position, but this approach is not energy efficient.

In this paper, we propose an efficient path-planning technique for USV and AUVs by which localization uncertainties are considered and the explored area is maximized. In the proposed method, AUVs form a chain of mobile beacons during exploration and improve localization accuracy through AUV-to-USV (A2U) and AUV-to-AUV (A2A) consensus communication. Each AUV should maintain a specific distance to its adjacent AUVs to be able to collaboratively inspect larger areas in an energy-efficient way.

### A. Literature Review and Our Contributions

The literature on AUVs addresses several major challenges, including the dynamic modeling and control of AUVs [11].

This work was supported by the Australian Research Council. Also, this work received funding from the Australian Government, via grant AUSMURIB000001 associated with ONR MURI grant N00014-19-1-2571.

M. Eskandari is with MAI OptiTek, Sydney, NSW, Australia, and was formerly with the School of Electrical Engineering & Telecommunications, UNSW, Sydney, NSW, Australia (e-mail:m.eskandari@maioptitek.com). A. Savkin is with the School of Electrical Engineering & Telecommunications, UNSW, Sydney, NSW, Australia (e-mail: a.savkin@unsw.edu.au). M. Deghat is with the School of Mechanical and Manufacturing Engineering, UNSW Sydney, NSW 2052, Australia (e-mail: m.deghat@unsw.edu.au).

Path tracking and formation control of a group of AUVs is the topic of the majority of research works [12], where various control methods such as robust and sliding mode control have been tested [13], [14]. AUV navigation, including path planning and localization, which is more relevant to this paper, is another research direction. Z. Zhou *et al.* discussed it within the context of cooperative navigation [15]. Bioinspired methods such as artificial fish/lily pad have been used for autonomous navigation [16]. Neural networks have been applied for AUV task assignment and path planning [17], [18]. Also, deep (reinforcement) learning-based methods are gaining popularity in underwater and robotic navigation [19], [20]. However, these methods have robustness issues, as deep learning and heuristic techniques require extensive data analysis and numerous trial-and-error iterations, which are difficult for underwater environments. Online view planning by creating a grid map of unexplored underwater structures is considered in [21], which is not practical for deep ocean exploration. Path planning is implemented for sensor data collection based on the value of information [22], [23]. The energy efficiency is considered through a hierarchical scheme including static-dynamic layers for data collection [24]. Also, sampling-based path planning is used for random trees-based trajectory design which causes time complexity issues in complex environments with large numbers of AUVs [25]. Vision-based path planning [26], and deep reinforcement learning (DRL) techniques are emerging topics [27]-[29]. Among the existing techniques, model predictive control (MPC) has proven to be a powerful technique and is becoming extremely popular for motion planning and tracking control for autonomous vehicles [30], [31].

The contributions of the paper are outlined as follows:
1) To the best of our knowledge, this paper is the first work that proposes a systematic localization-aware path-planning approach for an ocean-exploring team of autonomous vehicles including a USV and multiple AUVs. The proposed method is particularly useful for urgent marine search and rescue operations and scientific ocean exploration.
2) Path planning for the USV-AUVs team is formulated as a finite receding horizon MPC optimization problem, subject to AUV motion and nonholonomic constraints. A linear dynamic-aware kinematic equation of motion is developed for AUVs to be used in the path planning optimization algorithm.
3) The configuration of localization-aimed A2U and A2A consensus graphs is incorporated into the MPC optimization problem by defining proper objectives, as well as acoustic and line-of-sight (LoS) constraints. The time complexity of the optimization is elaborated indicating a nonconvex NP-hard problem.
4) A computationally tractable solution to the MPC optimization is proposed based on sequential convex programming. Appropriate convexification and decomposition techniques are proposed to ensure the solution is both feasible and optimal.
5) The feasibility and optimality of the method are analyzed using concepts from graph theory. Simulation results using MATLAB are provided to validate the proposed methodology.

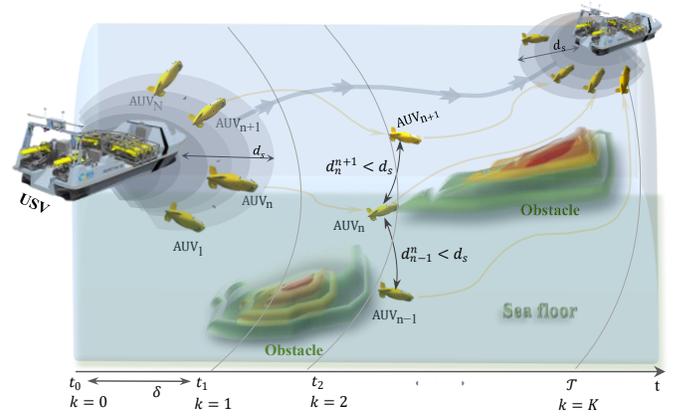

Fig. 1. Autonomous navigation of USV-AUVs for ocean exploration.

The remainder of this paper is organized as follows. In Section II, the problem description is presented followed by kinodynamic motion modeling. In Section III, the path planning is modeled as a finite horizon MPC optimization problem and its time complexity is discussed. In Section IV, the solution to the MPC optimization problem is presented along with a discussion on its feasibility and optimality. Simulations are conducted in Section V, and Section VI concludes the paper.

*Notations*: superscript ⊤ denotes matrix transpose; operator $\|.\|_2$ denotes the Euclidean norm, $\lceil.\rceil$ denotes the ceiling operator. Operator $\mathcal{L}_r(X_i, X_j)$ denotes linear interpolation between 3D coordinates $X_i$ and $X_j$ with a resolution $r$ in meters.

## II. PROBLEM DESCRIPTION AND MODELING

There are $N$ AUVs that are carried by the USV to a designated point in the ocean, from where AUVs depart from the USV for exploration and return once the mission is complete, see Fig. 1. After AUVs separate from the USV, they take routes away from the USV and other AUVs to maximize exploration coverage while keeping a minimum distance from the uneven sea floor, which may expose them to obstacles. We assume that the map of the sea floor is available as *a priori* information. The AUVs collect data and label it with time and localization information, i.e., position and orientation in the Earth's 3D coordinates. Due to the lack of GPS signals in the underwater environment, the AUVs should localize themselves using odometry commands and a dead reckoning system, utilizing INS [8]. However, this approach is vulnerable to error due to the odometry drift, particularly in the presence of uncertainties associated with the ocean environment (e.g., currents, discontinuities in temperature and salinity). Acoustic signalling is the only way of communication for correcting localization errors, but it is limited by the sonar range, multi-path issue, and background noise [1]. To handle these issues, we propose and solve the navigation problem as follows.

AUVs synchronize their clock to the global GPS time before departing from the USV and descending to the ocean floor. The path planning is implemented for the USV and AUVs by a finite horizon MPC optimizer, which considers the following issues:
1. Periodically, after the finite receding horizon with time interval ($\mathcal{T}$), all AUVs should be within the sonar range of the USV to communicate collected labeled data and remove the localization error.

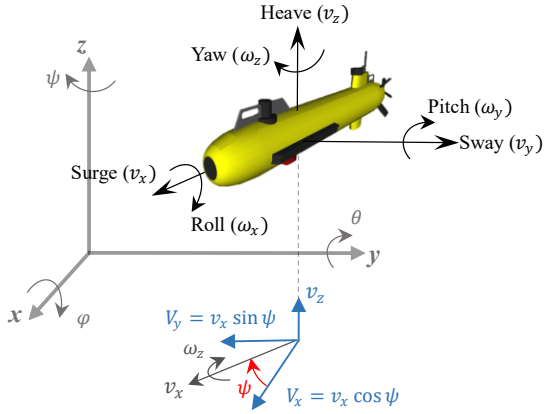

Fig. 2. AUV motion in the global 3D position-orientation $\{x, y, z, \varphi, \theta, \psi\}$ and body linear-angular velocity $\{v_x, v_y, v_z, \omega_x, \omega_y, \omega_z\}$ reference frames. The origin is located at the vehicle center of buoyancy.

2. The optimizer should generate collision-free paths close to the sea floor and with minimum energy effort to maximize the mission endurance and the safety of AUVs.
3. While navigating during time interval $\mathcal{T}$, for localization purposes, one AUV should be always within the sonar range ($d_s$) of the USV for A2U communication, and other AUVs should be in the sonar range of their adjacent AUVs for A2A consensus communication. The A2U and A2A graphs are optimally determined to allow AUVs to explore a larger area.
4. The localization status is updated by beaconing acoustic signals based on consensus protocol among the AUVs. Therefore, adjacent AUVs should keep moving in the LoS of each other for sonar communication.
5. Regarding the distribution of uncertainties and localization (dead-reckoning) errors, the path planning algorithm should ensure that each AUV remains close to the group to minimize the risk of them becoming lost.

The localization labels are captured by the AUV using its odometry and consensus-based beaconing localization, subject to background noise. Localization corrections can be performed using pose graph optimization regarding the known first and final positions as virtual loop closure. The probabilistic consensus protocol and pose graph optimizations are beyond the scope of this paper and will be addressed in future work.

### A. AUV Dynamic Motion Model

The AUV motion is modeled through the following 6 degrees of freedom (DoF) dynamic equation [32] that maps the linear-angular velocity of the AUV rigid body reference frame to the earth-fixed 3D (position and orientation) reference frame:

$$\dot{\eta} = \mathcal{J}(\eta)v; \quad (1)$$

$$\dot{v} = -\underbrace{M^{-1}(C(v) + D)}_{E(v)} v - G(\eta) + u + d; \quad (2)$$

where the notation of system variables is defined as follows. $\eta = [\eta_1, \eta_2]^\top \in \mathbb{R}^6$ with $\eta_1 = [x, y, z]^\top \in \mathbb{R}^3$ and $\eta_2 = [\varphi, \theta, \psi]^\top \in \mathbb{R}^3$ denote 3D position and orientation in the global (earth) 3D reference frame, respectively; $v = [v_1, v_2]^\top \in \mathbb{R}^6$ with $v_1 = [v_x, v_y, v_z]^\top \in \mathbb{R}^3$ and $v_2 = [\omega_x, \omega_y, \omega_z]^\top \in \mathbb{R}^3$ denote 3D linear (translational) and angular velocity in the AUV body reference frame, respectively; $u \in$ $\mathbb{R}^6$ denotes per mass/inertia control input vector and $d \in \mathbb{R}^6$ accounts for unmodeled dynamics and disturbances. The notation of system parameters is defined as follows, all of which are given in Appendix A. $\mathcal{J}(\eta) = \text{diag}\{\mathcal{J}_1(\eta_2), \mathcal{J}_2(\eta_2)\} \in \mathbb{R}^{6\times 6}$ denotes the time-varying Jacobian matrix where $\mathcal{J}_1(\eta_1) = \mathcal{R}_z(\psi)\mathcal{R}_y(\theta)\mathcal{R}_x(\varphi), \in \mathbb{R}^{3\times 3}$ is the 3D rotation matrix with $\mathcal{R}_x(\varphi)$, $\mathcal{R}_y(\theta)$, and $\mathcal{R}_z(\psi)$ as rotation matrices with respect to $x$, $y$, and $z$ axis [33], respectively, and $\mathcal{J}_2(\eta_2) \in \mathbb{R}^{3\times 3}$ denotes transformation of angular velocities; $M \in \mathbb{R}^{6\times 6}$, $C(v) \in \mathbb{R}^{6\times 6}$, $D \in \mathbb{R}^{6\times 6}$ represent the inertia, the Coriolis and centripetal mass forces, and hydrodynamic damping terms, respectively, and $G(\eta) \in \mathbb{R}^6$ combines per mass/inertia buoyancy and gravitational forces in the body reference frame.

Due to the prohibitive time complexity arising from the nonlinearity and large order of the model in path planning, a simplified model is required. Besides, the accurate values of the hydrodynamic matrices are not known and are difficult to obtain due to the uncertainties in the ocean environment [34]. However, the trajectory should satisfy the kino-dynamic constraints of the vehicle. Also, the odometry commands (which serve as the input to the propeller system) should be known for dead reckoning. Therefore, we use a probabilistic kinematic model.

### B. The Kinematic Motion Model

A preferably linear model is required to reduce the computational complexity of path planning, where some dynamic terms can be safely ignored in the AUV dynamic motion model in Fig. 2. In the horizontal plane, the roll rotational velocity can be neglected, i.e., $\varphi = 0$. Also, the AUV vertical motion is controlled by linear heave velocity, and the angular pitch velocity can be ignored, i.e., $\theta = 0$. Therefore, (1) is simplified as:

$$\begin{bmatrix} \dot{\eta}_1 \\ \dot{\psi} \end{bmatrix} = \begin{bmatrix} \mathcal{R}_z(\psi) & 0_{3\times 1} \\ 0_{1\times 3} & 1 \end{bmatrix} \begin{bmatrix} v_1 \\ \omega_z \end{bmatrix}. \quad (3)$$

Further, in the horizontal plane, the linear surge velocity is dominated compared to the sway velocity and the AUV heading can be controlled by the yaw rotational velocity. Therefore, the sway velocity is neglected, and (3) is simplified as:

$$\begin{bmatrix} \dot{\eta}_1 \\ \dot{\psi} \end{bmatrix} = \begin{bmatrix} \mathcal{R}'_z(\psi) & 0_{3\times 1} \\ 0_{1\times 2} & 1 \end{bmatrix} v', \quad (4)$$

where $\mathcal{R}'_z(\psi) = \begin{bmatrix} \cos\psi & \sin\psi & 0 \\ 0 & 0 & 1 \end{bmatrix}^\top \in \mathbb{R}^{3\times 2}$, and $v' = [v_x, v_z, \omega_z]^\top$. Also, the velocity dynamic in (2) is simplified as

$$\dot{v}' = -E'v' + u' + d', \quad (5)$$

where $E' = \text{diag}\{\gamma_{v_x}/(m - \gamma_{\dot{v}_x}), \gamma_{v_z}/(m - \gamma_{\dot{v}_z}), \gamma_{\omega_z}/J_z\} \in \mathbb{R}^{3\times 3}$; with $m$ denotes the AUV mass; $J_z$ denotes rotational inertia with respect to $z$ axis; $\gamma_{\dot{v}_x}, \gamma_{\dot{v}_z}, \gamma_{v_x}, \gamma_{v_z}$, and $\gamma_{\omega_z}$ are hydrodynamic parameters; $u' \in \mathbb{R}^3$ denotes modified input effort and $d' \in \mathbb{R}^3$ models lumped unmodeled dynamics, buoyancy gravitational forces, disturbances, etc. By stacking (4)-(5) we obtain

$$\begin{bmatrix} \dot{\eta}_1 \\ \dot{\psi} \\ \dot{v}' \end{bmatrix} = \begin{bmatrix} \mathcal{R}'_z(\psi) & 0_{3\times 1} \\ 0_{1\times 2} & 1 \\ -E' \end{bmatrix} v' + \begin{bmatrix} 0_{4\times 3} \\ I_{3\times 3} \end{bmatrix} (u' + d'). \quad (6)$$

where $I_{3\times3} \in \mathbb{R}^{3\times3}$ is the identity matrix. To handle the nonlinearity of $\mathcal{R}'_z(\psi)$ in (6), we omit angular velocity from the state space model and define new horizontal velocity variables:

$$\begin{cases} V_x = v_x \cos\psi \\ V_y = v_x \sin\psi \\ V_z = v_z \end{cases}, \quad (7)$$

where $v_x = \|V_x, V_y\|_2$, and $\psi = \tan^{-1}(V_y/V_x)$. Further, (6) is updated using (7) and is discretized as a linear discrete-time state-space model. In the following, the time interval of navigation $\mathcal{T}$ is discretized into a set of time steps $k \in \mathcal{K} = \{1, 2, \ldots, K\}$ with sampling time $\delta$ where $K = \lceil \mathcal{T}/\delta \rceil$. The subscript '$n$' is denoted as the $n^{th}$ AUV, where $n \in \mathcal{G} = \{1, 2, \ldots, N\}$, with $N$ is the number of AUVs in the AUV group $\mathcal{G}$. The discrete model is obtained as:

$$\underbrace{\begin{bmatrix} P_n[k+1] \\ V_n[k+1] \end{bmatrix}}_{X_n[k+1]} = \underbrace{\begin{bmatrix} I_{3\times3} & \delta I_{3\times3} \\ 0_{3\times3} & I_{3\times3} - \delta E'_n \end{bmatrix}}_{A_n} \underbrace{\begin{bmatrix} P_n[k] \\ V_n[k] \end{bmatrix}}_{X_n[k]} + \underbrace{\begin{bmatrix} 0_{3\times3} \\ I_{3\times3} \end{bmatrix}}_{B} \mathcal{U}_n[k] + \begin{bmatrix} I_{3\times3} \\ 0_{3\times3} \end{bmatrix} \varpi, \quad (8)$$

where $[k]$ denotes time step $k$; $P_n = [x_n, y_n, z_n]^\top \in \mathbb{R}^3$ denotes the position of the $n^{th}$ AUV in the 3D Euclidean space of the global (earth) reference frame; $V_n = \left[V_{x_n}, V_{y_n}, V_{z_n}\right]^\top \in \mathbb{R}^3$ denotes the 3D Euclidean linear velocity vector; $A_n$ and $B_n$ denote state and input matrices, respectively; $\mathcal{U}_n[k] \in \mathbb{R}^3$ denotes the input energy effort, and $\varpi = \mathcal{N}(0, \sigma^2) \in \mathbb{R}^3$ denotes Gaussian noise with zero mean and covariance $\sigma^2$ for modeling disturbances, and drift in the dead reckoning process. For path tracking purposes, the surge velocity input at sampling time $k$ is recovered as the 2-norm of the horizontal velocity vector at the corresponding sampling time, i.e., $v_{x_n}[k] = \left\|\left[V_{x_n}[k], V_{y_n}[k]\right]^\top\right\|_2$ and the AUV horizontal heading (i.e., angular velocity) is obtained as $\psi[k] = \tan^{-1}\left(V_{y_n}[k]/V_{x_n}[k]\right)$. The latter adds nonconvexity to the navigation which we will tackle with an appropriate solution by Lemma IV.1.

The full or augmented dynamic model can be used to follow the path in each sampling interval [32]. We put constraints corresponding to the AUV linear velocity and heading into the path planner to make the produced path feasible for path tracking. The heading change from $X_n[k-1] \to X_n[k]$ is given as:

$$\overrightarrow{\psi_n}[k] = \frac{1}{\delta}\left(\tan^{-1}\left(\frac{V_{y_n}[k]}{V_{x_n}[k]}\right) - \tan^{-1}\left(\frac{V_{y_n}[k-1]}{V_{x_n}[k-1]}\right)\right), \quad (8)$$

and will be considered a constraint to make the generated path nonholonomic. The following linear kinematic equation of motion is considered for the USV in 2D space:

$$\underbrace{\begin{bmatrix} x_{USV}[k+1] \\ y_{USV}[k+1] \end{bmatrix}}_{P_{USV}[k+1]} = \underbrace{I_{2\times2}}_{A_{USV}} \underbrace{\begin{bmatrix} x_{USV}[k] \\ y_{USV}[k] \end{bmatrix}}_{P_{USV}[k]} + \delta \underbrace{I_{2\times2}}_{B_{USV}} \underbrace{\begin{bmatrix} V_{x_{USV}}[k] \\ V_{y_{USV}}[k] \end{bmatrix}}_{\mathcal{U}_{USV}[k]}, \quad (9)$$

where $P_{USV}[k] \in \mathbb{R}^2$ denotes USV's 2D position in the global Euclidean space; $\mathcal{U}_{USV}[k] \in \mathbb{R}^2$ denotes USV's 2D velocity as the control input, and $\psi_{USV}[k] = \tan^{-1}\left(V_{y_{USV}}[k]/V_{x_{USV}}[k]\right)$ denotes USV heading at time step $k$.

## III. MPC-BASED PATH-PLANNING

The path-planning algorithm is designed in the context of finite receding horizon MPC optimization.

### A. The MPC Optimization Problem

The following MPC optimization problem ($\mathcal{P}1$) is considered:

$\mathcal{P}1$:

OF1. $\displaystyle\min_{\substack{\mathcal{U}[k] \,\forall k \in \mathcal{K} \\ \sigma_{i,j}\,\forall\, i,j\in\mathcal{G} \\ \sigma_{n,USV}\,\forall\, n\in\mathcal{G}}} \sum_{k=1}^{K} \Bigg( w_1 \underbrace{\|\mathcal{U}[k]\|_2^2}_{OF1_1} + w_2 \underbrace{\mathfrak{Z}\mathcal{X}[k]}_{OF1_2}$

$+ w_3 \underbrace{\sum_n \sigma_{n,USV}[k] \Big|\|(\mathfrak{P}_n - \mathfrak{P}_{USV})\mathcal{X}[k]\|_2 - d_s\Big|}_{OF1_3}$

$+ w_4 \underbrace{\sum_i^N \sum_{j>i}^N \sigma_{i,j}[k] \Big|\|(\mathfrak{P}_i - \mathfrak{P}_j)\mathcal{X}[k]\|_2 - d_s\Big|}_{OF1_4}$

$- w_5 \underbrace{\sum_i^N \sum_{j>i}^N (1-\sigma_{i,j}[k]) \|(\mathfrak{P}_i - \mathfrak{P}_j)\mathcal{X}[k]\|_2^2}_{OF1_5}$

$+ w_6 \underbrace{\sum_{i=1}^N \sum_{j>i}^N \sigma_{i,j}[k]}_{OF1_6} + w_7 \underbrace{\sum_n^N \sigma_{n,USV}[k]}_{OF1_7}$

$- w_8 \underbrace{\|\mathfrak{P}_{USV}\mathcal{X}[K] - \mathfrak{P}_{USV}\mathcal{X}[0]\|_2^2}_{OF1_8}\Bigg),$

s.t.

$C1_1.$ $\mathcal{X}[k+1] = \mathcal{A}\mathcal{X}[k] + \mathcal{B}\mathcal{U}[k],$ $\forall\, k\in\mathcal{K}$;

$C1_2.$ $\mathcal{X}[k+1] = \mathcal{A}^{k+1}\mathcal{X}[0] + \mathfrak{B}[k]\mathbb{U}[k],$ $\forall\, k\in\mathcal{K}$;

$C1_3.$ $\mathcal{X}[1:K] = \mathbb{A}\mathcal{X}[0] + \mathbb{B}\mathbb{U}[K],$

$C1_4.$ $\mathfrak{Z}_n\mathcal{X}[k] < \varepsilon,$ $\forall\, n\in\mathcal{G}, \forall\, k\in\mathcal{K}$;

$C1_5.$ $\|(\mathfrak{P}_n - \mathfrak{P}_{USV})\mathcal{X}[k]\|_2 \leq d_{max},$ $\forall\, n\in\mathcal{G}, \forall\, k\in\mathcal{K}$;

$C1_6.$ $\|(\mathfrak{P}_n - \mathfrak{P}_{USV})\mathcal{X}[k]\|_2 \leq d_s,$ $\forall\, n\in\mathcal{G}, \forall\, k=0,K$;

$C1_7.$ $\left\|\left([\mathcal{V}_n\mathcal{X}[k]]^\top\right)\right\|_2 \leq V_{n_{max}},$ $\forall\, n\in\mathcal{G}, \forall\, k\in\mathcal{K}$;

$C1_8.$ $\sigma_{n,USV}[k] \in \{0,1\},$ $\forall\, n\in\mathcal{G}, \forall\, k\in\mathcal{K}$;

$C1_9.$ $\sigma_{i,j}[k] \in \{0,1\},$ $\forall\, i,j>i\in\mathcal{G}, \forall\, k\in\mathcal{K}$;

$C1_{10}.$ $\sum_n^N \sigma_{n,USV}[k] \geq 1,$ $\forall\, k\in\mathcal{K}$;

$C1_{11}.$ $\sum_{i=1}^N \sum_{j>i}^N \sigma_{i,j}[k] \geq N-1,$ $\forall\, k\in\mathcal{K}$;

$C1_{12}.$ $\mathfrak{P}_n\mathcal{X}[k] \cap p_\Omega^{3D} = \emptyset,$ $\forall\, n\in\mathcal{G}, \forall\, k\in\mathcal{K}$;

$C1_{13}.$ $\sigma_{i,j}\,\mathcal{L}_r\left(\mathfrak{P}_i\mathcal{X}[k], \mathfrak{P}_j\mathcal{X}[k]\right) \cap p_\Omega^{3D} = \emptyset,$

$\forall\, i, j>i\in\mathcal{G}, \forall\, k\in\mathcal{K}$;

$C1_{14}.$ $\overrightarrow{\psi_n}[k] \leq \overrightarrow{\psi}_{max},$ $\forall\, n\in\mathcal{G}, \forall\, k\in\mathcal{K}$;

where
$$\mathcal{X}[k] = \left[P_{USV}^\top[k], X_1^\top[k], X_2^\top[k], \ldots, X_N^\top[k]\right]^\top \in \mathbb{R}^{(2+6N)};$$

$$\mathfrak{U}[k] = \left[\mathcal{U}_{USV}^\top[k], \mathcal{U}_1^\top[k], \mathcal{U}_2^\top[k], \ldots, \mathcal{U}_N^\top[k]\right]^\top \in \mathbb{R}^{(2+3N)};$$

$$\mathbb{U}[k] = \left[\mathfrak{U}^\top[0], \mathfrak{U}^\top[1], \mathfrak{U}^\top[2], \ldots, \mathfrak{U}^\top[k]\right]^\top \in \mathbb{R}^{(k+1)(2+3N)};$$

$$\mathcal{A} = \begin{bmatrix} A_{USV} & 0_{2\times 6} & \cdots & 0_{2\times 6} \\ 0_{6\times 2} & A_1 & & 0_{6\times 6} \\ \vdots & & \ddots & \vdots \\ 0_{6\times 2} & 0_{6\times 6} & \cdots & A_N \end{bmatrix} \in \mathbb{R}^{(2+6N)\times(2+6N)};$$

$$\mathbb{A} = [\mathcal{A}^1, \mathcal{A}^2, \ldots, \mathcal{A}^K]^\top \in \mathbb{R}^{K(2+6N)\times(2+6N)};$$

$$\mathcal{B} = \begin{bmatrix} B_{USV} & 0_{2\times 3} & \cdots & 0_{2\times 3} \\ 0_{6\times 2} & B_1 & & 0_{6\times 3} \\ \vdots & & \ddots & \vdots \\ 0_{6\times 2} & 0_{6\times 3} & \cdots & B_N \end{bmatrix} \in \mathbb{R}^{(2+6N)\times(2+3N)};$$

$$\mathfrak{B}[k] = [\mathcal{A}^k\mathcal{B}, \mathcal{A}^{k-1}\mathcal{B}, \ldots, \mathcal{A}\mathcal{B}, \mathcal{B}]^\top \in \mathbb{R}^{(2+6N)(k+1)\times(2+3N)};$$

$$\mathbb{B} = \begin{bmatrix} \mathcal{B} & 0 & & 0 & 0 \\ \mathcal{A}\mathcal{B} & \mathcal{B} & \cdots & 0 & 0 \\ \mathcal{A}^2\mathcal{B} & \mathcal{A}\mathcal{B} & & 0 & 0 \\ \vdots & & \ddots & & \vdots \\ \mathcal{A}^{K-1}\mathcal{B} & \mathcal{A}^{K-2}\mathcal{B} & \cdots & \mathcal{A}\mathcal{B} & \mathcal{B} \end{bmatrix} \in \mathbb{R}^{K(2+6N)\times K(2+3N)}$$

$$\mathfrak{Z} = [0_{1\times 2}, \mathfrak{x}_1, \mathfrak{x}_2, \ldots, \mathfrak{x}_N] \in \mathbb{R}^{1\times(2+6N)} \text{ with } \mathfrak{x}_n = [0_{1\times 2}, 1, 0_{1\times 3}];$$

$$\mathfrak{Z}_n = [0_{1\times 2}, 0_{1\times 6_1}, \ldots, \mathfrak{x}_n, \ldots, 0_{1\times 6_N}] \in \mathbb{R}^{1\times(2+6N)};$$

$$\mathfrak{P} = [0_{3\times 2}, \mathfrak{p}_1, \mathfrak{p}_2, \ldots, \mathfrak{p}_N] \in \mathbb{R}^{3\times(2+6N)} \text{ with } \mathfrak{p}_n = [I_{3\times 3}, 0_{3\times 3}];$$

$$\mathfrak{P}_n = [0_{3\times 2}, 0_{3\times 6_1}, \ldots, \mathfrak{p}_n, \ldots, 0_{3\times 6_N}] \in \mathbb{R}^{3\times(2+6N)};$$

$$\mathfrak{P}_{USV} = \begin{bmatrix} I_{2\times 2} \\ 0_{1\times 2} \end{bmatrix}, 0_{3\times 6_1}, \ldots, 0_{3\times 6_N} \in \mathbb{R}^{3\times(2+6N)};$$

$$\mathcal{V} = [0_{3\times 2}, \mathfrak{v}_1, \mathfrak{v}_2, \ldots, \mathfrak{v}_N] \in \mathbb{R}^{3\times(2+6N)} \text{ with } \mathfrak{p}_n = [0_{3\times 3}, I_{3\times 3}];$$

$$\mathcal{V}_n = [0_{3\times 2}, 0_{3\times 6_1}, \ldots, \mathfrak{v}_n, \ldots, 0_{3\times 6_N}] \in \mathbb{R}^{3\times(2+6N)};$$

$w_{1\sim 8}$ are weighting coefficients to scale problem objectives; $\sigma_{n,USV} \forall n \in \mathcal{G}$ and $\sigma_{i,j} \forall i, j \in \mathcal{G}$ are binary 0/1 variables; $d_s$ denotes the sonar modem range, $d_{max}$ denotes the maximum allowable distance between AUVs and the USV, $V_{n_{max}}$ denotes the maximum achievable velocity of AUV $n$, $p_\Omega^{3D}$ denotes a set of 3D coordinates of the sea floor obstacle that is supposed to be known and available, and $\emptyset$ denotes the empty set.

Objective functions (OF) $\text{OF1}_{1\sim 8}$ follow the following aims. $\text{OF1}_1$ minimizes the input energy effort for navigation; $\text{OF1}_2$ enforces the AUVs to navigate close to the sea floor for feature and data extraction. If there are hot spots or sensor networks, etc., that must be visited by the AUVs, this objective can be modified as the Euclidean distance between the AUVs and sensors; $\text{OF1}_3$ ensures that the AUVs selected by the binary variables $\sigma_{n,USV}, \forall n \in \mathcal{G}$ (in the A2U graph), remain within the sonar range of the USV for A2U communication and localization during navigating; $\text{OF1}_4$ makes the AUVs of the A2A graph be in the sonar range of each other, selected by binary variables $\sigma_{i,j}, \forall i, j \in \mathcal{G}$, for A2A consensus communication and localization during navigating; $\text{OF1}_5$ keeps non-adjacent AUVs to be apart for exploring larger areas; $\text{OF1}_{6,7}$ minimizes the number of adjacency constraints (to be in the sonar range for A2U and A2A communication) to allow the group to expand and explore a wider area, and $\text{OF1}_8$ enforces the USV to move away from the start point to allow AUVs to explore more. This objective function can be replaced with $\text{OF1}_{8\_2}$ as

$$\text{OF1}_{8\_2} = w_{8\_2} \left\| \mathfrak{P}_{USV} \mathcal{X}[K] - P_{USV}[\infty] \right\|_2^2;$$

where $P_{USV}[\infty]$ is the destination (or desired direction) toward which the exploration mission is carried out.

Constraints $\text{C1}_{1\sim 14}$ satisfy the following conditions. $\text{C1}_1$ governs the kinodynamic transition between consecutive time steps; $\text{C1}_2$ denotes finite receding horizon MPC that presents the kinodynamic transition from the initial point to time step $k+1$; $\text{C1}_3$ denotes the finite receding horizon MPC from the initial point to time step $K$ that presents all waypoints (states) in its output; $\text{C1}_4$ enforces the AUVs to stay underwater; $\text{C1}_5$ keeps the AUVs at the maximum allowable distance from the USV; $\text{C1}_6$ ensures all AUVs are within the sonar range of the USV at the first (start point) and last time steps for communication and localization (removing dead reckoning drift); $\text{C1}_7$ limits the AUVs to their maximum velocity; $\text{C1}_{8,9}$ specify $\sigma_{n,USV}$ and $\sigma_{i,j}, \forall i, j, n \in \mathcal{G}$ as binary variables; $\text{C1}_{10}$ ensures at least one AUV stays in the range of the USV (A2U graph) while navigating, $\text{C1}_{11}$ ensures the A2A adjacency graph among AUVs is complete and all AUVs can communicate through distributed consensus protocol; $\text{C1}_{12}$ satisfies collision-free (to the sea floor obstacles) navigation for crash avoidance; $\text{C1}_{13}$ enforces adjacent AUVs have LoS view for acoustic communication to minimize the multi-path effect, and $\text{C1}_{14}$ satisfies nonholonomic constraints.

*B. Time Complexity of the Optimization Problem*

$\mathcal{P}1$ is a quadratically constrained quadratic programming. The time complexity of the solution, e.g., with the interior-point method, is $\mathcal{O}(\mathfrak{M}^{1/2}(\mathfrak{M}+\mathfrak{N})\mathfrak{N}^2)$, where $\mathfrak{M}$ linearly increases based on the number of objectives and constraints, and $\mathfrak{N}$ denotes the number of variables and constants depending on sampling steps $K$ and AUV number $N$. However, $\mathcal{P}1$ is a nonconvex NP-hard optimization problem for the following reasons. $\text{OF1}_{3\sim 7}$ along with $\text{C1}_{8\sim 11}$ are mixed-integer programming, which is an NP-complete. But $\text{C1}_{12}$ and $\text{C1}_{13}$ make the $\mathcal{P}1$ an NP-hard as it is not possible to model the linear interpolation operator $\mathcal{L}_r$ in the convex programming as stated in Lemma III.1. Besides, $\text{OF1}_{3\sim 4}$ and $\text{C1}_{14}$ add nonconvexity to the optimization problem, which is another source of NP-hardness [35].

**Lemma III.1.** Energy-efficient obstacle-free path planning on uneven terrain/floors with valid LoS paths in the A2A graph is an NP-hard nonconvex optimization problem.

**Proof of Lemma III.1.** Known obstacles in 2D can be modeled as linear/quadratic constraints that convert the path planning to a mixed integer linear programming [36] that increases the time complexity by $\mathcal{O}(2^n)$, $n$ being the number of obstacles. However, it can be solved in exponential time with branch and bound method. In 3D, the obstacle-free minimum path planning problem is nonconvex due to the non-convexity of the uneven terrain/floor. Spots, where LoS links are not valid, can be modeled as virtual obstacles that should be avoided. The LoS paths can be checked by linear interpolation search with the time complexity of $\mathcal{O}(\mathbb{I})$, where $\mathbb{I}$ is associated with the resolution of the linear interpolation search. However, an infinite number of paths should be checked, which makes the problem formulation NP-hard in a nonconvex 3D space.

□

**Remark III.1.** The time complexity for path planning with a sampling-based method [37], e.g., rapidly exploring random trees (RRT), is impacted by the number of AUVs, sampling steps, and problem objectives/constraints and can be approximated by $\mathcal{O}\big((\mathfrak{MIN})N!\log((\mathfrak{MIN})N!)\big)$. The running time of the sampling methods imposes a prohibitive computational burden when the number of AUVs and sampling steps are large and is not feasible to solve it in polynomial time. Therefore, it is more efficient and practically feasible to tackle the nonconvexity and NP-harness of $\mathcal{P}1$ and solve it with convex optimization techniques.

## IV. THE SOLUTION TO THE MPC OPTIMIZATION

We solve $\mathcal{P}1$ using sequential convex optimization in three Steps. $\mathcal{P}1$ includes two sub-objectives: 1) collision-free energy-efficient path planning, and 2) A2U and A2A adjacency graph optimization including binary variables for A2A LoS acoustic communication and localization. First, in Step 1, we ignore the binary variables that specify A2U and A2A adjacency graphs, relax objectives and constraints associated with them, and solve $\mathcal{P}2$ to initialize energy-efficient obstacle-free paths. Then adjacency graphs are optimized by solving $\mathcal{P}3$ and $\mathcal{P}4$ in Step 2. Step 3 updates the paths by solving the MPC optimization with known binary variables obtained from Step 2.

### A. Sequential Convex Programming

**Step 1.** $\mathcal{P}1$ is modified in $\mathcal{P}2$ to initialize paths without A2U and A2A graph optimization and binary constraints.

$\mathcal{P}2$:

OF2. $\min\limits_{\mathbb{U}[k]\ \forall k\in\mathcal{K}} \sum\limits_{k=1}^{K}(w_1 OF1_1 + w_2 OF1_2 - w_8 OF1_8)$

s.t.

C2$_1$. C1$_{1\sim 7}$;

C2$_2$. $\mathfrak{Z}_n(\mathcal{A}^{k+1}\mathcal{X}[0] + \mathfrak{B}[k]\mathbb{U}[k]) \geq \mathbb{Z}_n[k] + \varepsilon,$
$\forall n \in \mathcal{G}, \forall k \in \mathcal{K};$

C2$_3$. $\begin{bmatrix}V_{x_n}[k]\\V_{y_n}[k]\end{bmatrix} - \left(1 + \frac{\alpha}{V_{h_{max}}}\right)\begin{bmatrix}V_{x_n}[k-1]\\V_{y_n}[k-1]\end{bmatrix} < 0$
$\forall n \in \mathcal{G}, \forall k \in \mathcal{K};$

C2$_4$. $\begin{bmatrix}V_{x_n}[k]\\V_{y_n}[k]\end{bmatrix} - \left(1 - \frac{\alpha}{V_{h_{max}}}\right)\begin{bmatrix}V_{x_n}[k-1]\\V_{y_n}[k-1]\end{bmatrix} > 0$
$\forall n \in \mathcal{G}, \forall k \in \mathcal{K};$

where $\alpha$ is a hyperparameter that controls the rate of change of horizontal linear speed. C2$_1$ repeats constraints C1$_{1\sim 7}$ associated with $\mathcal{P}1$; C2$_2$ models C1$_{12}$ and prevents collision with the sea floor with $\mathbb{Z}_n[k]$ indicating the depth of the sea, i.e., the $z-$coordinates of the sea floor corresponding to the horizontal $xy-$coordinates of the $n^{th}$ AUV at time step $k$, which is a varying parameter to model the uneven sea floor. We solve $\mathcal{P}2$ in an iterative manner where after each iteration, $\mathbb{Z}_n[k]$ is found and bounded. Based on the simulation results, $\mathcal{P}2$ converges after a few iterations with a linear time complexity of $\mathcal{O}(KN)$ and there is no computation issue associated with this constraint. C2$_{3,4}$ impose nonholonomic (heading change) constraints by linearizing C1$_{14}$ as per Lemma IV.1, given in the next subsection. Constraints C1$_{8\sim 11}$ and C1$_{13}$ are relaxed in $\mathcal{P}2$.

$\mathcal{P}2$ is convex and can be solved using quadratic optimization and gradient-based interior-point method in polynomial time. The solution to $\mathcal{P}2$ yields the energy-efficient paths for the USV-AUVs pack. Then, the problem is to expand the A2U and A2A graphs so that the exploration area is maximized, which is implemented in Step 2.

**Step 2.** Given $\mathbb{U}[k], \forall k = K$ from the solution to $\mathcal{P}2$, the A2U adjacency graph is determined in the initial path by solving $\mathcal{P}3$, to obtain $\sigma_{n,USV}[k], \forall k \in \mathcal{K}$ that represent A2U graph.

$\mathcal{P}3$ ($\forall k \in \mathcal{K}$):

OF3. $\min\limits_{\substack{\sigma_{n,USV}[k]\\ \forall n \in \mathcal{G}}} \sum\limits_{n}^{N}\Big(\sigma_{n,USV}[k] \times$
$\Big\|(\mathfrak{P}_n - \mathfrak{P}_{USV})\underbrace{(\mathcal{A}^{k+1}\mathcal{X}[0] + \mathfrak{B}[k]\mathbb{U}[k])}_{from\ \textbf{Step 1.}}\Big\|_2^2\Big)$

s.t.

C3$_1$. $0 \leq \sigma_{n,USV}[k] \leq 1,$ $\forall n \in \mathcal{G};$

C3$_2$. $\sum_n^N \sigma_{n,USV}[k] = 1,$

Since, $\mathbb{U}[k], \forall k = K$, is known, $\mathcal{P}3$ is linear and its solution can be obtained using the simplex method. At each sampling time, it selects the nearest AUV to the USV as the connection point of the A2U graph. Then, the A2A adjacency graph is determined by solving $\mathcal{P}4$ to obtain binary variables $\sigma_{i,j}[k]$, $\forall i, j \in \mathcal{G}, \forall k \in \mathcal{K}$, that represents A2A graph optimization.

$\mathcal{P}4$ ($\forall k \in \mathcal{K}$):

OF4. $\min\limits_{\substack{\sigma_{i,j}[k]\\ \forall i,j \in \mathcal{G}}} \sum\limits_{i}^{N}\sum\limits_{j>i}^{N}\Big(-\sigma_{i,j}[k]$
$\times \Big\|(\mathfrak{P}_i - \mathfrak{P}_j)\underbrace{(\mathcal{A}^{k+1}\mathcal{X}[0] + \mathfrak{B}[k]\mathbb{U}[k])}_{from\ \textbf{Step 1.}}\Big\|_2^2\Big)$

s.t.

C4$_1$. $0 \leq \sigma_{i,j}[k] \leq 1,$ $\forall i, j > i \in \mathcal{G};$

C4$_2$. $\sum_{i=1}^N \sum_{j>i}^N \sigma_{i,j}[k] = N - 1,$

C4$_3$. $1 \leq \sum_{j=1, j\neq i}^N \sigma_{i,j}[k] \leq 2,$ $\forall i \in \mathcal{G};$

where C4$_{1,2}$ keep $\sigma_{i,j}[k], \forall i,j > i \in \mathcal{G}$ as binary 0/1 variables; C4$_2$ ensures that the A2A graph is a connected graph, and C4$_3$ ensures that every AUV is connected to a minimum of one and a maximum of two AUVs so the maximum graph expansion is achieved. The goal in $\mathcal{P}4$ is to expand the A2A graph so that the AUVs pack explores wider areas while the A2A communication is satisfied. The solution to $\mathcal{P}4$, which is solved for each sampling time, selects edges for the A2A graph (i.e., $\sigma_{i,j}, \forall i,j > i \in \mathcal{G}$) and puts the corresponding edge length (distance) to be close to $d_s$ (see $OF1_{4,5}$). Therefore, we defined the objective in $\mathcal{P}4$ to find the longest simple path (i.e., tree) in the A2A graph as it results in more expansion in the A2A graph.

**Algorithm 1.**

1. **Upload** 3D sea floor map: $Z(X,Y), \forall X, Y \in$ exploring area
2. $\text{for}_1\ \mathcal{T} = 1, 2, \ldots$
3.    $\text{if}_1\ \mathcal{T} == 1$,
4.       $P_{USV}[0] = [0\ \ 0]^\top$;
5.       $P_n[0] = rand(1,3)\ |\ \|P_{USV}[0] - P_n[0]\|_2 \leq d_s, \forall n$;
6.    **else**
7.       $P_{USV}[0] = P_{USV}[K], P_n[K], K \in \mathcal{T} - 1$;
8.    $\text{end if}_1$
9.    **Obtain** $\mathbb{U}[k]\ \forall\ k = K$ by solving $\mathcal{P}2$ (**Step 1**);
10.       $\text{if}_2\ z_n[k] \leq Z(x_n, y_n), \forall\ n \in \mathcal{G}, \forall\ k \in \mathcal{K}$
11.          $z_n[k] = Z(x_n, y_n) + \varepsilon$;
12.       $\text{end if}_2$
13.    $\text{for}_2\ k \in \mathcal{K}$
      Using $\mathbb{U}[k]$:
14.       **Determine** A2U graph by solving $\mathcal{P}3$ (**Step 2**)
      (Obtain $\sigma_{n,USV}[k], \forall n \in \mathcal{G}$)
15.       **Determine** A2A graph by solving $\mathcal{P}4$ (**Step 2**)
      (Obtain $\sigma_{i,j}[k], \forall ni, j \in \mathcal{G}$)
16.    $\text{end for}_2$
17.    **Solve** $\mathcal{P}5$ to obtain $\mathbb{U}[K]: \{\mathcal{U}[k],\ \forall k \in \mathcal{K}\}$; (**Step 3**)
18.    $\text{for}_3\ k \in \mathcal{K}$
19.       $\text{if}_3\ \sigma_{i,j}\ \mathcal{L}_r(\mathfrak{P}_i\mathcal{X}[k], \mathfrak{P}_j\mathcal{X}[k]) \cap p_\Omega^{3D} \neq \emptyset$,
20.          $d_{s,i,j} = d_s - \varepsilon$;
21.       $\text{end if}_3$
22.    $\text{end for}_3$
23. $\text{end for}_1$

Feasibility and optimality are discussed in the next subsection, where more discussion on $\mathcal{P}4$ and its objective constraints are provided.

**Step 3.** After obtaining the binary variables the original MPC optimization problem $\mathcal{P}1$ is converted to $\mathcal{P}5$ and is solved to obtain MPC input commands.

$\mathcal{P}5$:

$$\text{OF5.} \min_{\mathcal{U}[k]\ \forall k \in \mathcal{K}}\ \text{OF2} + \sum_{k=1}^{K} w_5 \text{OF1}_5$$

s.t.

$\text{C5}_1.\quad \text{C2}_{1\sim 4}$,

$\text{C5}_2.\quad d_s - d_t < \left(\underbrace{\sigma_{n,USV}[k]}_{from\ \text{Step 2}} \|(\mathfrak{P}_n - \mathfrak{P}_{USV})\mathcal{X}[k]\|_2^2\right) < d_s,$
$$\forall\ k \in \mathcal{K};$$

$\text{C5}_3.\quad d_s - d_t < \left(\underbrace{\sigma_{i,j}[k]}_{from\ \text{Step 2}} \|(\mathfrak{P}_i - \mathfrak{P}_j)\mathcal{X}[k]\|_2^2\right) < d_s,$
$$\forall\ i, j > i \in \mathcal{G}, \forall\ k \in \mathcal{K};$$

where $d_t$ denotes tolerance in the sonar range $d_s$.

In this step, $\text{OF1}_{3,4}$ are modeled by $\text{C5}_{2,3}$, and $\text{OF1}_{6,7}$ are given in Step 2. All other objectives/constraints have been modeled except for $\text{C1}_{13}$ which is realized using the following simple technique.

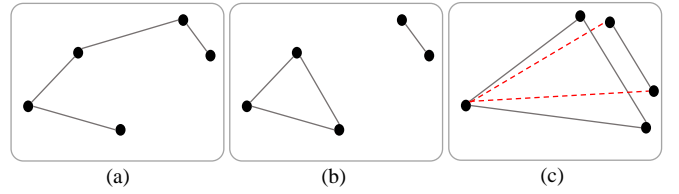

Fig. 3. Graph definitions: (a) a tree in the graph; (b) node clustering in the shortest tree algorithm; (c) contradiction of node clustering in the longest tree algorithm (solid lines indicate a possible solution in the longest path algorithm with $\mathcal{M} = 2$ node clusters; dashed lines indicate $\mathcal{M} = 2$ longer edge exists)

After solving $\mathcal{P}5$, $\text{C1}_{13}$ is checked for the A2A graph at each time step to validate LoS paths in a linear time complexity of $\mathcal{O}(K(N-1)\mathbb{I})$. If LoS is not valid for an A2A path, $\text{C5}_{2,3}$ corresponding to that edge are updated by reducing the value of $d_s$ associated with that edge. This technique along $\text{C5}_2$ adjusts the nodes of the corresponding A2A edge to ensure that LoS becomes valid.

The proposed MPC path planning optimization problem is summarized in Algorithm 1.

*B. On the Feasibility and Optimality of the Solution*

In Step 1, the nonconvexity of heading control is tackled using the convexity relaxation presented in Lemma IV.1.

**Lemma IV.1.** The rate of heading change of a vehicle can be controlled by bounding the lower bound and upper bound of the rate of change of linear horizontal speeds in the x and y directions.

**Proof of Lemma IV.1.** In the horizontal plane the vehicle's velocity is controlled by controlling linear velocities, i.e., $V_x$ and $V_y$ in x and y directions, respectively. The heading angle is obtained as $\psi = \tan^{-1}(V_y/V_x)$. The rate change of the heading is derived as

$$d\psi = \frac{V_x}{V_x^2 + V_y^2} dV_y - \frac{V_y}{V_x^2 + V_y^2} dV_x. \tag{10}$$

In discrete time, by limiting the lower/upper bounds of $\Delta V_x = V_x[k] - V_x[k-1]$ and $\Delta V_y = V_y[k] - V_y[k-1]$, $\Delta \psi$ can be linearly bounded. Assuming that the largest possible value for linear horizontal speed is $V_{h_{max}}$, the upper bounds of $V_x[k]$ and $V_y[k]$ can be constrained by limiting the difference between the consecutive slots as

$$\begin{cases} V_x[k] - V_x[k-1] > \dfrac{aV_x[k-1]}{V_{h_{max}}} \\ V_y[k] - V_y[k-1] > \dfrac{aV_y[k-1]}{V_{h_{max}}} \end{cases} \tag{11}$$

as given in $\text{C2}_3$. With a similar statement, $\text{C2}_4$ limits the lower bound of $\Delta V_x$ and $\Delta V_y$ so that $\text{C2}_{3,4}$ limit rapid heading change.
□

In Step 2, the A2U and A2A graphs are optimized by solving $\mathcal{P}3$ and $\mathcal{P}4$ for each sampling time. It is not necessary to preserve the same A2U and A2A graph topologies throughout the navigation (i.e., for all $k's$). Therefore, flexible graph topologies are considered for A2U and A2A communication. This allows the pack to freely converge to the optimal path for exploring the ocean environment. Nevertheless, optimizing the A2A graph requires more attention, which is highlighted in Lemma IV.2, after the following definition of a tree.

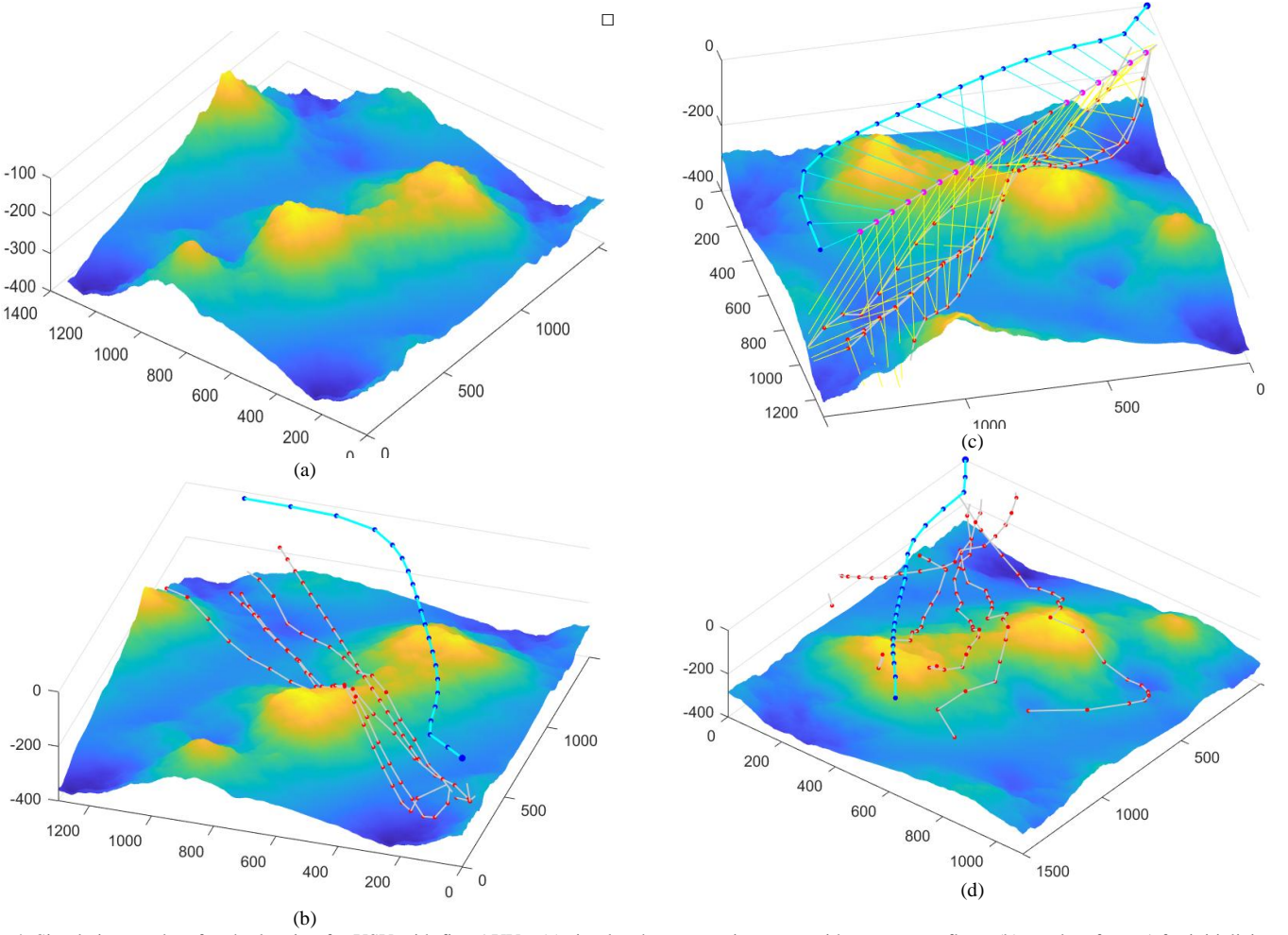

Fig. 4. Simulation results of path planning for USV with five AUVs: (a) simulated ocean environment with uneven sea floor; (b) results of step 1 for initializing energy-efficient paths with floor constraints; (b) results of step 2 for A2U and A2A graph optimization; (c) results of step 3 for final paths.

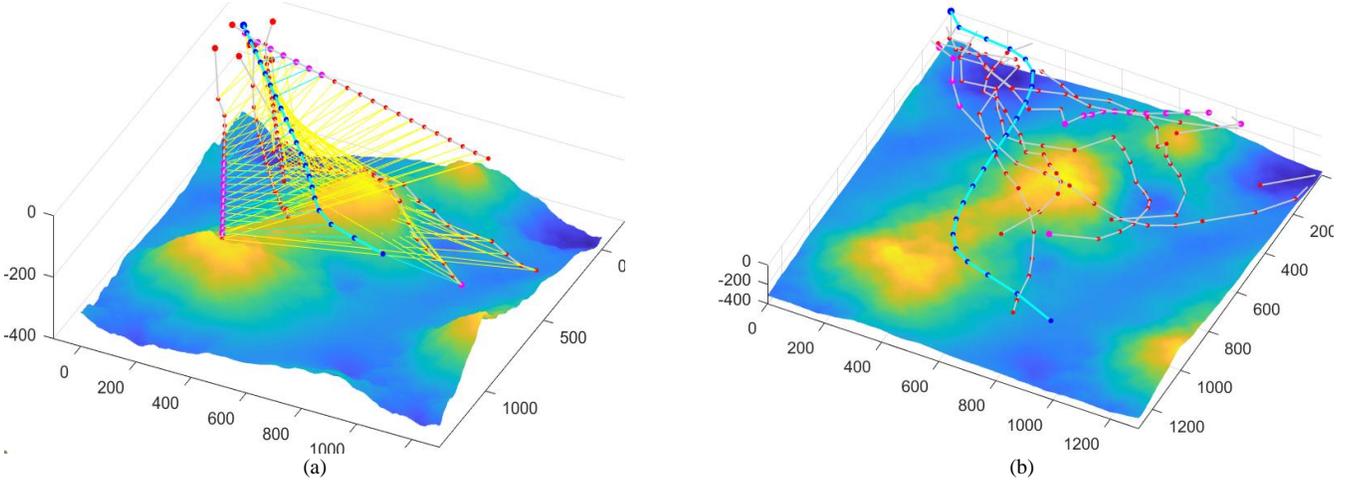

Fig. 5. Simulation results of path planning for USV with seven AUVs; (a) results of step 1 and step 2 for initializing energy-efficient paths and A2U and A2A graph optimization; (b) results of step 3 for final paths.

**Definition IV.1.** A *tree* in the graph is defined as a simple path that passes through each node only once without forming any cycle, as illustrated in Fig. 3(a).

**Lemma IV.2.** The A2A graph out of $\mathcal{P}4$ should span the *longest tree* among AUV nodes to preserve the optimality and feasibility of the MPC convex programming for generating the minimum energy path.

**Proof of Lemma IV.2.** From the definition of a tree, it is obvious that the longest tree is the largest tree in a graph that spans $N-1$ edges, where $N$ is the number of nodes in the graph, see Fig 3(a). Therefore, $C4_2$ is defensible, which indicates that $\mathcal{P}4$ spans a tree. Also, $C4_3$ ensures that the obtained A2A graph is a simple tree that passes each node once.

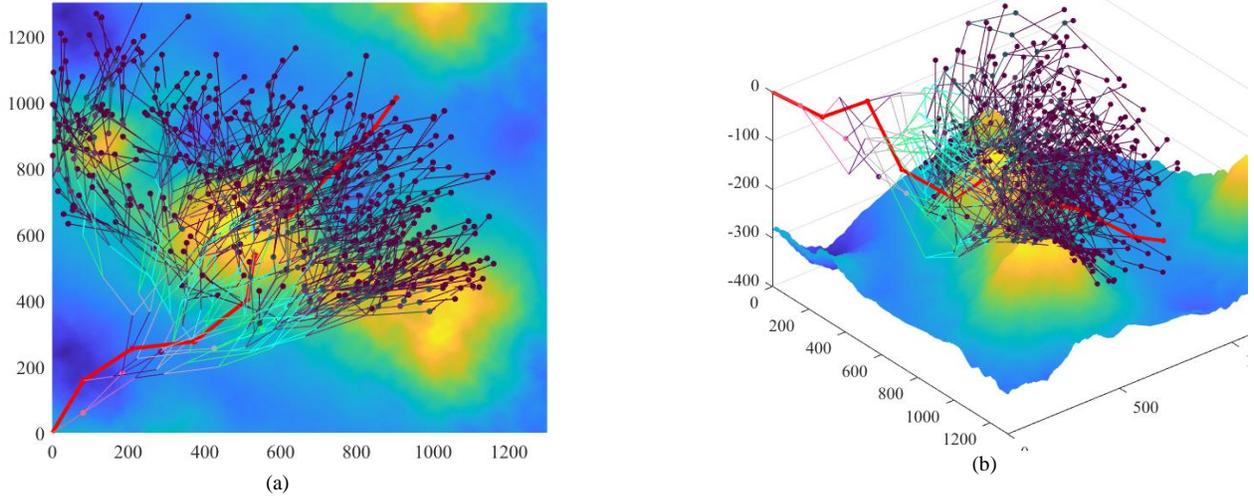

Fig. 6. Simulation results of path planning with the RRT-like method; (a) 2D view; (b) 3D view.

However, the no-cycling condition of a tree cannot be satisfied by $C4_{1\sim3}$ as these constraints may trap in node clustering, shown in Fig. 3(b) if they tend to span the shortest path. The node clustering issue is problematic for A2A consensus graph as AUVs may divide into two or more groups with a distance larger than $d_s$. Besides, spanning the shortest tree in a graph is known as the traveling salesman problem, which is an NP-hard problem and makes convex programming infeasible in polynomial time. Nevertheless, achieving the longest tree, as designed in OF4, prevents the node clustering issue, see Lemma IV.3. Further, it preserves the MPC optimality for developing the energy-efficient path as follows. Consider the following objective function:

$$\mathcal{P}6: \quad \min_{\sigma_{i,j}, \forall i,j \in \mathcal{G}} \sum_{1}^{N-1} (d_s - \sigma_{i,j} d_{ij}),$$

s.t. $C4_1, C4_2,$ and $C4_3,$

where $d_{ij} < d_s \ll d_{max}$, and $d_{ij}$ is the distance between node $i$ and node $j$ (i.e., AUV $i$ and AUV $j$) given from the initial energy-efficient path out of Step 1. $d_{ij} < d_s$ is a valid assumption as the MPC optimizer for the energy-efficient path tends to keep AUVs close to each other. $d_{ij}$ can be expanded to $d_s$ if $\sigma_{i,j} = 1$. Therefore, the larger $d_{ij}$, the less effort is required to make it $d_s$ in the A2A consensus graph. □

**Lemma IV.3.** Spanning the longest tree by $\mathcal{P}4$ does not result in node clustering.
**Proof of Lemma IV.3.** We prove this by contradiction. Suppose that the longest tree algorithm in $\mathcal{P}4$ results in node clustering with $\mathcal{M}$ clusters, as shown by the solid lines in Fig. 3(c). By inspection, there are still $\mathcal{M}$ longer edges than the currently spanned edges that can reorder the spanned graph to a longer tree. This would cause the algorithm to converge to the longest tree without node clustering.

## V. SIMULATION RESULTS

The proposed MPC optimization path planning technique is simulated in the MATLAB platform. The MPC optimization is solved for the receding 40 minutes including 20 time slots with a sampling time of 100 seconds. In the first optimization, it is assumed that the AUVs have departed from the USV and are in random positions, but in the sonar range of the USV, $ds = 150$ m. At the next MPC optimization for the receding 40 minutes, the initial positions of the AUVs and the USV are set based on their final positions from the preceding MPC optimization. Without loss of generality, we assume the first position of the USV is at the origin and the AUVs are randomly distributed in the half-underwater sphere with radius $d_s$. Also, $C1_6$ enforces the AUVs to approach the USV at the final time step for communication and localization.

The proposed MPC optimization problem is formulated by MATLAB coding and developing numerical matrices with appropriate sizes as presented in $\mathcal{P}1$. Then, it is solved with convex quadratic programming utilizing the nonlinear optimization toolbox in MATLAB. Simulation results of path planning with MPC optimization for a USV with five AUVs are presented in Fig. 4. Fig. 4(a) shows the ocean environment simulated with uneven seafloor including underwater valleys and seamounts to simulate obstacle-free navigation. Fig. 4(b) illustrates the initialized obstacle-free energy-efficient paths while considering the uneven seafloor. The optimization problem in Step 1 successfully initializes paths that are close to the seafloor and $C2_2$ prevents collisions with obstacles. The A2U and A2A graph optimization results in Step 2 are shown in Fig. 4(c). At each step time, the nearest AUV to the USV is selected and the A2U graph is represented by light blue lines. Also, the A2A graph is determined in this step and the results are shown with yellow lines. The final paths are shown in Fig. 4(d) where AUVs are distributed over the seafloor for exploration, but they tend to converge toward the mother USV at final time steps. The running time of solving the optimization problems is captured with the MATLAB *tic toc* command, where Step 1 ($\mathcal{P}2$), Step 2 ($\mathcal{P}3 + \mathcal{P}4$), and Step 3 ($\mathcal{P}5$) were solved in 10.1, 0.13 + 0.26, and 51.2 seconds, respectively.

Fig. 5 illustrates the simulation results for a USV with seven AUVs. Fig. 5(b) indicates that the proposed technique properly controls the heading change and the USV path is nonholonomic. Running times for Steps 1 to 3 were recorded as 23.05, 0.13 + 0.57, and 117.19 seconds, respectively.

To evaluate the effectiveness of the proposed technique, in terms of running time and real-time path planning for autonomous navigation, we simulated a sampling-based method for a single AUV. Since the endpoint is uncertain, we implemented the following RRT-like path-planning approach to generate random trees [38]. For every time step $k = 1:K$, $\mathfrak{r}$ valid random samples are generated for the AUV waypoints. To validate these random samples, problem constraints, such as obstacle-free and motion constraints (speed, heading, etc.,) are checked. A random tree that achieves the most optimal scores in terms of energy ($OF1_1$), depth ($OF1_2$), and convergence ($OF1_{8\_2}$) indices is selected as the final path. The time complexity of the sampling method, when following the proposed optimization process for calculating objective functions and graph structure, can be approximated by

$$\mathcal{O}\left(\underbrace{(\mathfrak{M}\mathcal{NS})\log(\mathfrak{M}\mathcal{NS})}_{\text{Path objectives and constraints}} + \underbrace{(2N)^{K-1}}_{\text{A2U \& A2A graphs}}\right), \quad (12)$$

where $\mathcal{S} = \sum_k^K \mathfrak{r}^k$ is the total number of random samples and trees, and $\mathfrak{M}$ linearly increases based on the number of objectives and constraints. Simulation results are shown in Fig. 6. The running time for only a single AUV is more than 300 seconds which will be significantly higher with more AUVs and random waypoints for more optimal solutions. It reveals that sampling is prohibitive for real-time path planning for a team of AUVs.

## VI. Discussion

The proposed path planning method for a team of AUVs is advantageous from several perspectives: 1) The performance of the AUV team is improved for high-quality ocean exploration with efficiency in time and energy. 2) AUVs can benefit from other AUVs in the A2U and A2A consensus graph where AUVs act as a chain and localization error can be removed using a beaconing consensus protocol. 3) The proposed method systematically facilitates inception, including localization and path planning for autonomous navigation. 4) The obtained MPC commands can be used for path tracking and thus would simplify the formation control. 5) Last but not least, the problem can be solved in real-time, thanks to the developed linear motion model, the proposed MPC optimization problem as well as the convexification and decomposition techniques for the solution. The MPC optimization was formulated as a finite horizon path planning, by which it is not necessary to solve the optimization at each sample time. Besides, the path planner has sufficient time to solve the optimization problem of the receding path while the team follows the preceding path.

However, there are some issues with the proposed method that can be considered for future research extensions: 1) the AUV motion model is simplified for path planning. Utilizing motion primitives generated by path planning can be problematic for an underactuated AUV. However, they can be used by a path smoothing technique to generate motion primitives for sway and pitch motions. 2) The performance of the MPC depends on the weighting coefficients. The DRL technique can be used to improve the performance of the proposed MPC optimization by tuning the hyperparameters. The DRL agent could include graph neural networks (GNN) for modeling consensus graphs. This integration addresses practical challenges in real-time ocean exploration missions, which may not be fully addressed by AI-based methods alone.

## VII. Conclusion

In this paper, the autonomous navigation of a team of AUVs, coordinated by a USV, was studied for deep ocean exploration. It was stated that autonomous navigation requires an integrated localization-aware, yet computationally tractable path-planning algorithm. The path planning was formulated as a finite horizon MPC optimization with A2U and A2A consensus graphs for acoustic communication and localization. The time complexity of the optimization problem was discussed, and effective solutions were proposed to be solved in real time. The feasibility and optimality of the path planner algorithm were analytically verified based on sequential convex programming and graph theory concepts. Finally, simulation results in the MATLAB platform, including comparative results with sampling-based methods, demonstrated the efficacy and superiority of the proposed path-planning technique.

Future research could explore the integration of DRL techniques to improve overall performance and computation time. Additionally, GNN can be employed to model consensus graphs in the DRL agent more effectively.

## Appendix A

The rotation matrices are given as follows.

$$\mathcal{R}_x(\varphi) = \begin{bmatrix} 1 & 0 & 0 \\ 0 & \cos\varphi & -\sin\varphi \\ 0 & \sin\varphi & \cos\varphi \end{bmatrix}; \mathcal{R}_y(\theta) = \begin{bmatrix} \cos\theta & 0 & \sin\theta \\ 0 & 1 & 0 \\ -\sin\theta & 0 & \cos\theta \end{bmatrix};$$

$$\mathcal{R}_z(\psi) = \begin{bmatrix} \cos\psi & -\sin\psi & 0 \\ \sin\psi & \cos\psi & 0 \\ 0 & 0 & 1 \end{bmatrix};$$

$$\mathcal{J}_1(\eta_2) = \begin{bmatrix} c\,\theta\,c\,\psi & s\,\varphi\,s\,\theta\,c\,\psi - c\,\varphi\,s\,\psi & s\,\varphi\,s\,\psi + c\,\varphi\,s\,\theta\,c\,\psi \\ c\,\theta\,s\,\psi & s\,\varphi\,s\,\theta\,s\,\psi + c\,\varphi\,c\,\psi & c\,\varphi\,s\,\theta\,s\,\psi - s\,\varphi\,c\,\psi \\ -s\,\theta & s\,\varphi\,c\,\theta & c\,\varphi\,c\,\theta \end{bmatrix}$$

where $s$ and $c$ represent $\sin(\cdot)$ and $\cos(\cdot)$ functions, respectively.

$$\mathcal{J}_2(\eta_2) = \begin{bmatrix} 1 & \sin\varphi\tan\theta & \cos\varphi\tan\theta \\ 0 & \cos\varphi & -\sin\varphi \\ 0 & \sin\varphi/\cos\theta & \cos\varphi/\cos\theta \end{bmatrix};$$

$$M = \text{diag}\left\{m - \gamma_{\dot{v}_x}, m - \gamma_{\dot{v}_y}, m - \gamma_{\dot{v}_z}, J_x, J_y, J_z\right\} \in \mathbb{R}^{6\times 6},$$

$$D = -\text{diag}\left\{\gamma_{v_x}, \gamma_{v_y}, \gamma_{v_z}, \gamma_{\omega_x}, \gamma_{\omega_y}, \gamma_{\omega_z}\right\} \in \mathbb{R}^{6\times 6},$$

where $m$ denotes the AUV mass; $J_x, J_y$, and $J_z$ denote rotational inertia with respect to $x$, $y$, and $z$ axis, respectively; $\gamma_{\dot{v}_x}, \gamma_{\dot{v}_y}, \gamma_{\dot{v}_z}, \gamma_{v_x}, \gamma_{v_y}, \gamma_{v_z}, \gamma_{\omega_x}, \gamma_{\omega_y}, \gamma_{\omega_z}$ are hydrodynamic parameters. $C(v) = C_R(v) + C_A(v)$;

$$C_R(v) = \begin{bmatrix} 0_{3\times 3} & C_{R1}(v) \\ C_{R1}(v) & C_{R2}(v) \end{bmatrix}; C_A(v) = \begin{bmatrix} 0_{3\times 3} & C_{A1}(v) \\ C_{A1}(v) & C_{A2}(v) \end{bmatrix};$$

$$C_{R1}(v) = m\begin{bmatrix} 0 & v_z & -v_y \\ -v_z & 0 & v_x \\ v_y & -v_x & 0 \end{bmatrix};$$

$$C_{R2}(v) = m\begin{bmatrix} 0 & J_z\omega_z & -J_y\omega_y \\ -J_z\omega_z & 0 & J_x\omega_x \\ J_y\omega_y & -J_x\omega_x & 0 \end{bmatrix};$$

$$C_{A1}(\boldsymbol{v}) = \begin{bmatrix} 0 & \gamma_{\dot{v}_z}v_z & -\gamma_{\dot{v}_y}v_y \\ -\gamma_{\dot{v}_z}v_z & 0 & \gamma_{\dot{v}_x}v_x \\ \gamma_{\dot{v}_y}v_y & -\gamma_{\dot{v}_x}v_x & 0 \end{bmatrix};$$

$$C_{A2}(\boldsymbol{v}) = \begin{bmatrix} 0 & \gamma_{\omega_z}\omega_z & -\gamma_{\omega_y}\omega_y \\ -\gamma_{\omega_z}\omega_z & 0 & \gamma_{\omega_x}\omega_x \\ \gamma_{\omega_y}\omega_y & -\gamma_{\omega_x}\omega_x & 0 \end{bmatrix};$$

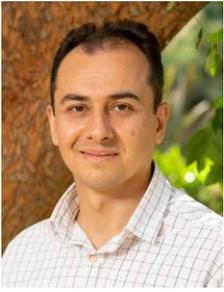
**Mohsen Eskandari** was born in Saveh, Iran. He received the Ph.D. in electrical engineering from the University of Technology Sydney (UTS), Sydney, Australia, in 2021. He is a Research Fellow at the University of New South Wales, Sydney, Australia. He has more than ten years of experience in different parts of the electrical industry. He has proven skills in handling power system projects as well as a strong background in the field of automation and control.

He is the founder of MAI OptiTek, where he provides technical AI-based consultancy services toward future intelligent grids and wireless networks. His research interests include power systems, future energy, AI-assisted control optimization and automation, wireless networks and robotics.

Dr. Eskandari is the recipient of the Higher Degree Research Excellence Award at UTS Faculty of Engineering and Information Technology in 2018 and 2019, in recognition of outstanding academic performance and the best individual HDR project.

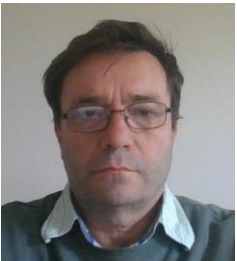
**Andrey V. Savkin** was born in Norilsk, USSR, in 1965. He received the M.S. and Ph.D. degrees in mathematics from Leningrad State University, Saint Petersburg, Russia, in 1987 and 1991, respectively. In 1987–1992, he worked with the All-Union Television Research Institute as an Engineer and as a Senior Research Scientist. In 1992–1994, he was a Postdoctoral Research Fellow with the Department of Electrical Engineering, Australian Defence Force Academy, UNSW, Canberra, ACT, Australia. In 1992–1994, he was a Research Fellow with the Department of Electrical Engineering, University of Melbourne, Melbourne, VIC, Australia. In 1996–2000, he was a Senior Lecturer and then an Associate Professor with the Department of Electrical and Electronic Engineering, University of Western Australia, Perth, WA, Australia. He has been a Professor with the School of Electrical Engineering and Telecommunications, University of New South Wales, Sydney, NSW, Australia, since 2000.

Prof. Savkin served as an Associate Editor/Editor for numerous international journals in the field. He has authored/coauthored nine research monograph and numerous journal papers in these areas. His research interests include control engineering, robotics, UAV navigation, power systems, wireless sensor networks, and biomedical engineering.

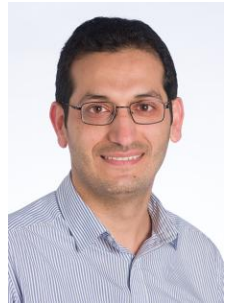
**Mohammad Deghat** (S'09–M'14) received the B.Sc. and M.Sc. degrees in electronics and electrical engineering from Shiraz University, Iran, and the Ph.D. degree in engineering and computer science from the Australian National University, Canberra, in 2013. He is a Lecturer (Assistant Professor) at the School of Mechanical and Manufacturing Engineering, University of New South Wales, Sydney. He was previously a Research Fellow at the University of Melbourne and had been a Researcher at the Commonwealth Scientific and Industrial Research Organisation (CSIRO) and NICTA. His research interests include multi-agent systems, sensor networks, localization and robot control.